# Spectral2Spectral: Image-spectral Similarity Assisted Deep Spectral CT Reconstruction without Reference

Xiaodong Guo, Yonghui Li, Dingyue Chang, Peng He, Peng Feng, Hengyong Yu, Weiwen Wu

*Abstract*—Spectral computed tomography based on a photon-counting detector (PCD) attracts more and more attentions since it has the capability to provide more accurate identification and quantitative analysis for biomedical materials. The limited number of photons within narrow energy bins leads to imaging results of low signal-noise ratio. The existing supervised deep reconstruction networks for CT reconstruction are difficult to address these challenges because it is usually impossible to acquire noise-free clinical images with clear structures as references. In this paper, we propose an iterative deep reconstruction network to synergize unsupervised method and data priors into a unified framework, named as Spectral2Spectral. Our Spectral2Spectral employs an unsupervised deep training strategy to obtain high-quality images from noisy data in an end-to-end fashion. The structural similarity prior within image-spectral domain is refined as a regularization term to further constrain the network training. The weights of neural network are automatically updated to capture image features and structures within the iterative process. Three large-scale preclinical datasets experiments demonstrate that the Spectral2spectral reconstructs better image quality than other the state-of-the-art methods.

## I. INTRODUCTION

The PCD-based spectral CT system has strong ability to identify an absorption property of scanned object within different energy ranges, demonstrating a great potential in contrast resolution enhancement, material identification, and quantitative analysis of biomedical soft tissues. PCD uses selectable energy thresholds to record x-ray photons of different energies, and the attenuation coefficients of scanned object can be reconstructed in these different energy bins. Different materials generally show different attenuation maps within different energy bins. Certainly, some materials have similar components and they may have similar intensities in reconstructed images within the same energy bin or different energy bins. It is difficult to distinguish them. However, a certain material has a specific attenuation coefficient within an energy bin. These different attenuation coefficients within different energy bins do not have a regular mathematical relationship due to the K-edge, and they are only decided by material itself and energy bins. To some extent, this characteristic can help to distinguish materials with similar components. However, there are many quantum noises in the acquired projections within narrow energy bins, resulting in lower signal-noise ratio (SNR) of reconstructed images. Therefore, how to reconstruct high-quality CT images from noisy projections is a challenge in the field of spectral CT.

In recent years, compressed sensing (CS)-based [1, 2] iterative reconstruction algorithms have been developed, which can reconstruct high-quality spectral CT images from noisy or incomplete projection data. For example, Zhang *et al*. applied a tensor dictionary learning (TDL) [3] technique to improve image quality of spectral CT reconstruction based on the similarity of images among different energy bins. Wu *et al*. introduced an image gradient $\ell_0$-norm into TDL to reconstruct low-dose spectral CT images, which is very effective to deal with image artifacts and noises [4]. In addition, other CS-based reconstruction methods were also developed for spectral CT, including [5], [6] and [7], *etc*.

Besides, there are some spectral CT reconstruction algorithms based on an iterative framework. Yu *et al*. proposed an iterative reconstruction algorithm named spectral prior image constrained compressed sensing (PICCS) [8]. He *et al*. developed a split-Bregman based iterative algorithm and combined a low-rank correlation descriptor with a structure extraction operator as a prior regularization term [9]. Some research utilized the similarity among different energy bins. Shi *et al*. combined a region-specific texture model with a low-rank correlation descriptor as a prior regularization [10]. Chen *et al*. applied a fourth-order nonlocal tensor decomposition model for spectral CT image reconstruction (FONT-SIR) [11]. Rigie *et al*. presented total nuclear variation (TVN) as a regularizer for reconstructing multi-channel, spectral CT images [12].

This work was supported in part by the National Natural Science Foundation of China (No: 62201628); in part by the Natural Science Foundation of Chongqing (No. CSTB2022NSCQ- MSX0360); in part by the National Key Research and Development Program of China (No: 2022YFA1204201); in part by the Guangdong Basic and Applied Basic Research Foundation (No: 2023A1515011780). (Corresponding authors: Peng He, Hengyong Yu and Weiwen Wu)
Contribution of Xiaodong Guo, Yonghui Li and Dingyue Chang are equal.
X. Guo, Y. Li, P. He and P. Feng are with the Key Lab of Optoelectronic Technology and Systems of the Education Ministry of China, Chongqing University, Chongqing, China( e-mail: 20165953@cqu.edu.cn, penghe@cqu.edu.cn, gxd@cqu.edu.cn, coe-fp@cqu.edu.cn).
D. Chang is with the China Academy of Engineering Physics, Institute of Materials, Mianyang, China (changdy2014@pku.edu.cn).
H. Yu is with the Department of Electrical and Computer Engineering, University of Massachusetts Lowell, Lowell, MA 01854, USA( e-mail: hengyong-yu@ieee.org ).
W. Wu is with the School of Biomedical Engineering, Sun Yat-Sen University, Shenzhen, Guangdong, China (e-mail: wuweiw7@mail.sysu.edu.cn). He is also with the Henan Key Laboratory of Imaging and Intelligent Processing, PLA Strategic Support Force Information Engineering University, Zhengzhou, China.

Currently, deep learning (DL) technology has been widely used in the fields of biomedical imaging and tomographic reconstruction [13]. In the post-processing tomographic reconstruction, Chen *et al*. [14, 15] proposed a low-dose CT reconstruction method based on deep learning. It is named residual encoder-decoder convolutional neural network (RED-CNN) and use normal dose dataset as label. FBPConvNet [16] used an U-Net [17] to improve image quality over raw FBP reconstructions. In addition, DD-Net [18], wavelet-transform-based U-net [19, 20], BCD-Net [21], and others [22] were further developed. Although these methods can obtain better results with good efficiency, they require the ground truths for network training. Regarding the end-to-end tomographic reconstruction, He *et al*. [23] designed a novel framework to approximate the Radon inversion. Zhu *et al*. [24] proposed a full manifold encoding–decoding convolutional architecture. However, these reconstruction methods require large amounts of paired data and high computing costs, particularly the large GPU memory. Since it is very difficult to obtain high-quality image for spectral CT, these deep reconstruction networks fail to achieve good reconstructed image quality. Hence, some deep reconstruction methods were developed to reconstruct spectral CT images without reference. Niu *et al*. presented a Noise2Sim unsupervised post-processing network [25]. Noise2Sim used the similarity from the neighboring slices as a prior knowledge to help network training. The main contribution is that Noise2Sim presents a new solution to construct pseudo-labels for spectral CT denoising. They choose adjacent slices as target image. Before training a network, dissimilar mask is used to remove dissimilar structure between input and target images. The processed target images play a pseudo-label role for network training. However, it is difficult to find two similar images for spectral CT, especially in fan-beam geometry. Besides, the reconstruction performance is also limited by the spatial similarity from two slices. Finally, it is one of the post-processing deep denoising networks, and its performance is limited by data consistency.

Inspired by the Noise2Sim, we propose a Spectral2Spectral unsupervised reconstruction network, which synergizes the model knowledge (i.e., simultaneous iterative reconstruction technique, SIRT) and data prior into a unified mathematical model. It reconstructs high-quality images from measurement data, and an unsupervised neural network is developed to optimize the reconstructed images. First, Spectral2Spectral incorporates a deep learning network into an iterative reconstruction framework to achieve a stable and reliable deep reconstruction solution. Second, the quality of reconstructed image is improved with an advanced unsupervised network by designing a novel loss function with the unique properties of spectral CT. Third, the proposed reconstruction model updates the weights automatically in the iterative process to adaptively optimize the reconstructed images.

The contributions of this study are threefold. First, we design a novel system to combine neural networks and conventional iterative reconstruction techniques within a regularization optimization framework. This strategic combination leverages the prior knowledge encoded within neural networks as a regularization constraint, ultimately enhancing image quality. Furthermore, we introduce a tailored loss function that quantifies image-spectral similarity within the neural network. It tangibly enhances image details and sharpens edges within the reconstructed images. Secondly, by introducing a deep neural network, we approximate intermediate iterative images, thereby enhancing the efficiency of the reconstruction process. The network's weighting parameters are adaptively updated with each iteration, facilitating the learning of adaptive image structures and features. Thirdly, we pioneer an unsupervised iterative learning approach for spectral CT reconstruction. Unlike conventional deep learning-based methods which demand meticulously labeled datasets, our approach sidesteps this requirement by generating pseudo-labels. These pseudo-labels are constructed directly from the spectral CT images themselves, diminishing the influence of noise in the reconstruction process.

The rest of the paper is organized as follows. In section II, we first briefly introduce the related work. In section III, we describe our proposed Spectral2spectral deep reconstruction network in details. In section IV, the specific experimental design, results and ablation studies are presented and analyzed. In the last section, we discuss some relevant issues and conclude the paper.

## II. RELATED WORK

### A. Unsupervised Denoising

Recently, convolutional neural networks (CNNs) provide powerful tools for image denoising. The Noise2Noise [26] was first developed by Lehtinen *et al*. to train a deep denoiser using several noisy observations of the same image. Subsequently, Noise2Void [27] was developed to use a masking technique for image denoising, where the neural network was employed to learn how to fill pixel gaps within the noisy image. Because this method only denoises the underlying picture, it's unable to learn the noise distribution. Although Noise2Void model was trained on a large image dataset with the same noise level, it may be adapted to denoise a single similar image without extra inputs and weights updating. Noise2Self [28] further enhanced and generalized this fundamental approach. Besides, the Self2Self [29] as an unsupervised technique was developed for single picture denoising, which can achieve competitive performance to the fully trained approaches (i.e., Noisy2Clean). Recorrupted2Recorrupted (R2R) [30] outperformed Self2Self on real-world noise by corrupting the input picture into fresh noisy realization pairs. Noise2Sim believes that an image usually has multiple self-similarities/repeated places (self-similarities/repeated). Given a patch within an image, one can always globally search one similar patch to construct an image patch pair for noise removal.

Noise2Noise first demonstrated that neural network models can be trained only using noisy images. Here, we brief introduce the main idea. The Noise2Noise requires the data to meet two conditions: (a) each scene $s$ has at least two noisy images $x_1$ and $x_2$; (b) the noises within $x_1$ and $x_2$ must be independent and zero-mean. The training aim of Noise2Noise

optimization is to minimize the following loss

$$\min_{\theta} E_{s,x_1,x_2} \|f_\theta(\mathbf{x}_1) - \mathbf{x}_2\|_F^2, \quad (1)$$

where $f_\theta$ represents the training network and $\theta$ represents the network parameters. From the above given conditions, different noise level image-pairs are necessary for Noise2Noise training. However, such a strict condition is difficult to be satisfied, especially for spectral CT imaging.

### B. Deep Learning in Spectral CT Reconstruction

To the best of our knowledge, deep learning based spectral CT reconstruction is difficult since there is no way to obtain sufficient high-quality images as references to train supervised networks. To obtain high-quality images, we previous proposed an U-net, $L_p$-norm, Total variation, Residual learning, and Anisotropic adaption (ULTRA) network [31], which uses the weighted anisotropic total variation as the regularization term for network training. The ULTRA has achieved good results in numerical simulations and clinical trials. However, the training of ULTRA requires labeled images, which makes it difficult to be adopted in practice.

Furthermore, as one of the unsupervised deep learning techniques, the Noise2Sim attempts to denoise spectral CT images and already shows its great potential to address this task. The Noise2Sim borrowed the idea of Noise2Noise to construct pseudo-labels by searching similar image-pairs. In this work, the spectral CT images of adjacent slices are treated as similar images. Assume two noisy images $\mathbf{x}_1 = \mathbf{s} + \mathbf{n}_1$ and $\mathbf{x}_2 = \mathbf{s} + \mathbf{n}_2 + \boldsymbol{\delta}$, where $\boldsymbol{\delta}$ is the difference between clean images, $\mathbf{n}_1$ and $\mathbf{n}_2$ are two independent and zero-mean noise images. The parameters $\theta$ of the denoising network are optimized as:

$$\min_{\theta} E_s \|f_\theta(\mathbf{s} + \mathbf{n}_1) - (\mathbf{s} + \mathbf{n}_2 + \boldsymbol{\delta})\|_2^2. \quad (2)$$

The noise assumption of Noise2Sim can be described as $E[\mathbf{n}_2|\mathbf{n}_1 + \mathbf{s}] = E[\mathbf{n}_2|\mathbf{s}] = 0$. Through searching the similar image patches, $E[\boldsymbol{\delta}|\mathbf{n}_1 + \mathbf{s}] \approx 0$ can be ensured in practice. That is, Noise2Sim is approximately equivalent to Noise2Noise when the images are similar enough.

## III. SPECTRAL2SPECTRAL FRAMEWORK

### A. Overall Architecture

Fig.1(a) shows the proposed Spectral2Spectral framework, which is similar to the traditional iterative optimization reconstruction method. Regarding the network architecture, it can be divided into three modules: iteration framework (i.e., SIRT), novel unsupervised network design, and compositeloss function optimization. Specifically, the classical SIRT reconstruction technique is first performed on the input sinogram to reconstruct an initial image. Second, an unsupervised network is employed to approximate the intermediate reconstruction results.

Here, our unsupervised network is adaptively updated with iteration, which means our Spectral2Spectral network can reconstruct images to continuously approach real situations. To characterize the similarity and correlations across different energy bins, a novel loss function is proposed to enhance image feature recovery and detail fidelity. In fact, we can further clearly clarify our Spectral2Spectral with three basic theories: reconstruction optimized theory, unsupervised denoising theory, and regularization optimization theory.

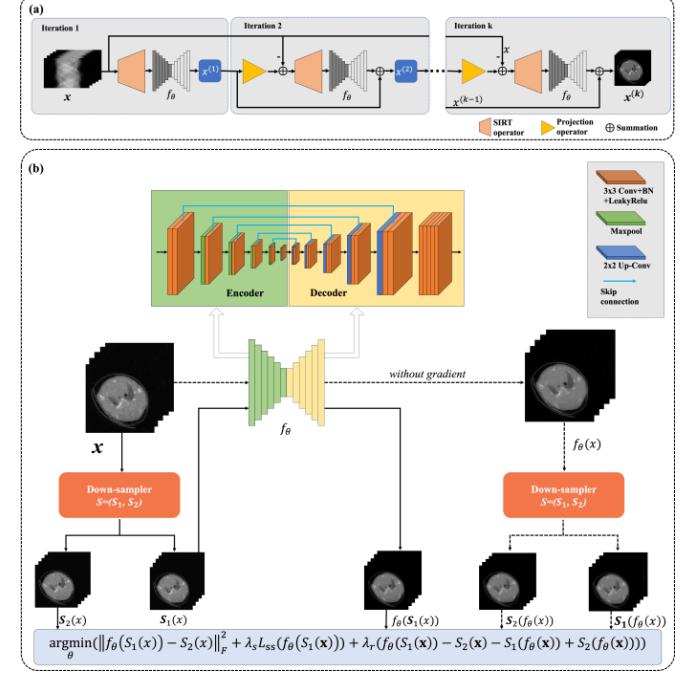

Fig. 1 The flowchart of Spcetral2Spectral. (a) is the workflow of the proposed method, and (b) is the training process.

### B. Optimized Reconstruction

The image reconstruction of a typical spectral CT system with cone-beam geometry can be formulated as:

$$\mathbf{y}_i = \mathbf{A}\mathbf{x}_i + \boldsymbol{\epsilon}_i, \quad i = 1, \ldots, I, \quad (3)$$

where $\mathbf{y}_i \in R^M$ represents the projection data from $i^{th}(1 \leq i \leq I)$ energy bin, and $I$ represents the number of energy channels. $M = N_e \times N_v$, where $N_e$ and $N_v$ are the numbers of the PCD elements and sampling view, respectively. $\boldsymbol{\epsilon}_i \in R^M$ represents noise within measurements, $\mathbf{x}_i \in R^J (J = H_1 \times H_2 \times H_3$ where $H_1$, $H_2$ and $H_3$ represent the size of reconstructed volumetric image), and $\mathbf{A} \in R^{M \times J}$ is the CT system matrix. Theoretically speaking, the solution can be achieved by operating the matrix inverse over Eq. (3). However, it is not applicable in practice since the system matrix is too huge to be implemented directly. Rather than to solve this problem directly, it can be iteratively handled by minimizing an optimization function

$$\min_{\{\mathbf{x}_i\}_{i=1}^I} \frac{1}{2} \sum_{i=1}^I \|\mathbf{y}_i - \mathbf{A}\mathbf{x}_i\|_F^2, \quad (4)$$

where $\|\cdot\|_F^2$ represents Frobenius norm. To stabilize the reconstruction results for ill-posed problems, a regularization term that characterizes prior knowledge can be incorporated into Eq. (4), and we have

$$\mathcal{X}^* = \underset{\mathcal{X}}{\operatorname{argmin}} \left( \frac{1}{2} \sum_{i=1}^I \|\mathbf{y}_i - \mathbf{A}\mathbf{x}_i\|_F^2 + \lambda \Phi(\mathcal{X}) \right), \quad (5)$$

where $\frac{1}{2}\sum_{i=1}^I \|\mathbf{y}_i - \mathbf{A}\mathbf{x}_i\|_F^2$ represents data consistency term, $\Phi$ represents the regularization term, $\lambda$ is a factor to balance the aforementioned two terms, and $\mathcal{X}$ represents one tensor formulated by rolling all energy-channel images $\{\mathbf{x}_i\}_{i=1}^I$. The CS-theory does not explicitly point out the specific sparse transformation, and it only requires that the observation matrix should meet the RIP (Restricted Isometry Property) conditions. That means there are many different possible sparse models based on the CS-theory. The regularization term is generally the most essential part of an optimization model. Different regularization priors can induce different image reconstruction algorithms. The most popular regularization priors include the total variation (TV) and its variants, the tight framelet, dictionary based sparse representation, nonlocal means, and low-rank matrix factorization [32-35]. The goal of Eq. (5) is to determine an optimal solution by minimizing the entire objective function. The split-Bregman method can be employed to solve this problem. Although the alternating direction method of multipliers (ADMM) can also solve this problem, the split-Bregman method has not the penalty parameter issue that requires an infinite value for an ideal solution. Under appropriate conditions, the split-Bregman method is equivalent to the ADMM algorithm [36]. Here, we first introduce a parameter $\mathcal{Z}$ to re-express Eq. (5) as follows

$$\{\mathcal{X}^*, \mathcal{Z}^*\} = \underset{\{\mathcal{X},\mathcal{Z}\}}{\mathrm{argmin}} \left( \frac{1}{2} \sum_{i=1}^{I} \|\mathbf{y}_i - \mathbf{A}\mathbf{x}_i\|_F^2 + \lambda \Phi(\mathcal{Z}) \right). \quad (6)$$
$$\text{subject to } \mathcal{Z} = \mathcal{X}.$$

This constrained optimization problem can be converted into an unconstrained optimization problem:

$$\{\mathcal{X}^*, \mathcal{Z}^*\} = \underset{\{\mathcal{X},\mathcal{Z}\}}{\mathrm{argmin}} \left( \begin{array}{c} \frac{1}{2} \sum_{i=1}^{I} \|\mathbf{y}_i - \mathbf{A}\mathbf{x}_i\|_F^2 + \\ \lambda \Phi(\mathcal{Z}) + \frac{\lambda_1}{2} \|\mathcal{Z} - \mathcal{X}\|_F^2 \end{array} \right). \quad (7)$$

Eq. (7) can be divided into two sub-problems:

$$\mathcal{X}^{(k+1)} = \underset{\mathcal{X}}{\mathrm{argmin}} \left( \frac{1}{2} \sum_{i=1}^{I} \|\mathbf{y}_i - \mathbf{A}\mathbf{x}_i\|_F^2 + \frac{\lambda_1}{2} \|\mathcal{X} - \mathcal{Z}^{(k)}\|_F^2 \right), (8)$$

$$\mathcal{Z}^{(k+1)} = \underset{\mathcal{Z}}{\mathrm{argmin}} \left( \frac{\lambda_1}{2} \|\mathcal{Z} - \mathcal{X}^{(k+1)}\|_F^2 + \lambda \Phi(\mathcal{Z}) \right), \quad (9)$$

where $k$ indicates the iteration number. Eq. (8) can be solved by operating the partial derivative of $\mathcal{X}$, that is

$$\mathbf{A}^T \sum_{i=1}^{I} (\mathbf{A}\mathbf{x}_i - \mathbf{y}_i) + \lambda_1 (\mathcal{X} - \mathcal{Z}^{(k)}) = 0, \quad (10)$$

where $\mathbf{A}^T$ represents the transpose of system matrix. Eq. (10) can be reformatted to:

$$(\mathbf{A}^T\mathbf{A} + \lambda_1)\mathcal{X} = (\mathbf{A}^T\mathbf{A} + \lambda_1)\mathcal{X}^{(k)} +$$
$$\mathbf{A}^T(\mathbf{y} - \mathbf{A}\mathcal{X}^{(k)}) + \lambda_1 (\mathcal{Z}^{(k)} - \mathcal{X}^{(k)}), \quad (11)$$

then $x$ can be obtained by an iterative method:

$$\mathcal{X}^{(k+1)} = \mathcal{X}^{(k)} +$$
$$(\mathbf{A}^T\mathbf{A} + \lambda_1)^{-1}\{\mathbf{A}^T(\mathbf{y} - \mathbf{A}\mathcal{X}^{(k)}) + \lambda_1 (\mathcal{Z}^{(k)} - \mathcal{X}^{(k)})\}. \quad (12)$$

In fact, Eq. (12) is the conventional iterative reconstruction formula with appropriate approximation to $(\mathbf{A}^T\mathbf{A} + \lambda_1)^{-1}$, which can be updated with SIRT reconstruction technique. Eq. (9) is a typical image restoration problem, and here one unsupervised convolutional neural network is proposed to adaptively approximate its solution. Based on the above theoretical analysis, adding deep learning to iterative reconstruction has a solid theoretical basis.

*C. Unsupervised Approximation Network*

For the image restoration problem Eq. (9), the neural network is an effective strategy for optimization. In fact, the intermediate raw image is usually tarnished by noises and artifacts in spectral CT. It can be generated from the joint distribution $p(\mathbf{x}_i, \mathbf{n}) = p(\mathbf{x}_i)p(\mathbf{n}|\mathbf{x}_i)$. The aim of neural network $f_\theta$ is to learn a mapping function to recover the clean image from noisy image, where $f$ is parameterized by the weights $\boldsymbol{\theta}$.

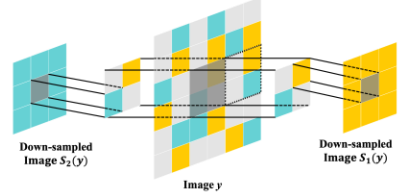

Fig. 2 Operation principle of down-sampler ($S_1(\cdot), S_2(\cdot)$)

The iterative reconstruction framework is usually fixed, and the final reconstruction of spectral CT often depends on the design of unsupervised denoising network. Our approximation network is inspired by a recently published approach called Neighbor2Neighbor [37] where the neural network learns to map adjacent pixels in the image to one-another, and the adjacent pixels tend to have a similar underlying signal. Here, we assume there are only small differences within a clean image pair: $s$ and $s + \delta$, and $\delta \to 0$. $x_1$ and $x_2$ are the noisy versions of $s$ and $s + \delta$, respectively. The following equation is established

$$E_{s,x_1}\|f_\theta(x_1) - s\|_F^2 =$$
$$E_{s,x_1,x_2}\|f_\theta(x_1) - x_2\|_F^2 - \sigma_{x_2}^2 + 2\delta E_{s,x_1}(f_\theta(x_1) - s). \quad (13)$$

Please refer to [37] for proof of Eq.(13). From Eq. (13), it can be found that when $\delta \to 0$, $2\delta E_{s,x_1}(f_\theta(x_1) - s) \to 0$. This means $(x_1, x_2)$ training pair can be used as an approximation of Noise2Noise. Therefore, if one finds the proper image pair $x_1$ and $x_2$ satisfying the property of "similar but not identical" ($\delta \to 0$), the denoising network can be trained. The essential problem is to construct such "similar but not identical" image pairs. One possible way of searching "similar but not identical" image pair for a single noisy image is down-sampling. The feasible sub-images can be sampled from adjacent but different positions of original noisy image. Such sub-images will satisfy the condition "similar but not identical" and ensure their difference is small and their corresponding clean images are not the same. Specifically, to sample a noisy image $\mathbf{x}$, a pair of nearest adjacent down-sampler $(S_1(\cdot), S_2(\cdot))$ are used to generate $(S_1(\mathbf{x}), S_2(\mathbf{x}))$. As shown in Fig.2, the image $y$ is divided into $(W_1/k) \times (W_2/k)$ patches. $W_1$ and $W_2$ are length and width of the image $y$. To illustrate this strategy, we set $k = 2$. For each patch, two neighboring locations are randomly selected. Then, they are treated as the pixels at the corresponding positions of $(S_1(\mathbf{x}), S_2(\mathbf{x}))$. After this step is repeated for all the patches, two sub-sampled images with size $(W_1/k) \times (W_2/k)$ are generated. To train a neural network in a Noise2Noise fashion, Eq. (1) can be rewritten as

$$\min E_{s,\mathbf{x}} \|f_\theta(S_1(x)) - S_2(x)\|_F^2. \quad (14)$$

Because the sampling positions of $S_1(\mathbf{x})$ and $S_2(\mathbf{x})$ are different, that is
$$E_{\mathbf{x}|\mathbf{s}}(S_1(\mathbf{x})) \neq E_{\mathbf{x}|\mathbf{s}}(S_2(\mathbf{x})). \quad (15)$$
If we only employ Eq. (14) to train the network directly, it is unlikely to yield optimal results but over-smoothed images. This situation can be corrected by adding a regularization term $L_r$ to the loss function. For an ideal denoise network, it should satisfy the following two conditions:
$$\begin{cases} f_\theta(\mathbf{x}) = \mathbf{s} \\ f_\theta(S_i(\mathbf{x})) = S_i(\mathbf{s}), i \in \{1,2\} \end{cases}. \quad (16)$$
Thus, we obtain the following expression
$$\begin{aligned} L_r &= E_{\mathbf{x}|\mathbf{s}}\{f_\theta(S_1(\mathbf{x})) - S_2(\mathbf{x}) - (S_1(f_\theta(\mathbf{x})) - S_2(f_\theta(\mathbf{x})))\} \\ &= S_1(\mathbf{s}) - E_{\mathbf{x}|\mathbf{s}}\{S_2(\mathbf{x})\} - (S_1(\mathbf{s}) - S_2(\mathbf{s})) \\ &= S_2(\mathbf{s}) - E_{\mathbf{x}|\mathbf{s}}\{S_2(\mathbf{x})\}, \end{aligned} \quad (17)$$
and $L_r$ can be considered as a regularization term for Eq. (14)
$$min(E_{s,x}\|f_\theta(S_1(x)) - S_2(x)\|_F^2 + \lambda_r L_r). \quad (18)$$
$L_r$ is used to prevent the denoising network from overfitting, and $\lambda_r > 0$ should gradually increase with the training progresses.

*D. Spectral Regularization Prior*

The loss function in Eq. (18) usually causes image being over-smoothed as well as finer structures losing, a new regularization term is introduced to recover blurry details. To further explore the sparse potential of spectral CT reconstruction, it is an effective way to incorporate image-spectral characteristic into the loss function. First, it is well-known that different energy channel images have different grey intensities, but they share the same image structures since channel-wise images are from the same physical object. It seems feasible to establish a unique loss to explore the relationship across image-spectral space. Empirically, the image from full-spectrum by adding the photons of all energy bins often provides a higher SNR than that obtained from a single energy channel. Therefore, the full-spectrum reconstructed image can be treated as the reference to calculate SSIM (structural similarity index measure) [38]. By considering the relationship within spectral-cross domain, we propose a structural similarity (SSIM) loss function $L_{ss}$ to characterize the similarity within image-spectral domain. Here, the SSIM can be given as
$$SSIM(\mathbf{x},\mathbf{y}) = \frac{(2\mu_\mathbf{x}\mu_\mathbf{y} + C_1)(2\sigma_{\mathbf{xy}} + C_2)}{(\mu_\mathbf{x}^2 + \mu_\mathbf{y}^2 + C_1)(\sigma_\mathbf{x}^2 + \sigma_\mathbf{y}^2 + C_2)}, \quad (19)$$
where $\mathbf{x}$ and $\mathbf{y}$ are two images, $\mu_\mathbf{x}$ and $\sigma_\mathbf{x}^2$ respectively represent the mean and the variance of $\mathbf{x}$, and $\sigma_{\mathbf{xy}}$ is the covariance of $\mathbf{x}$ and $\mathbf{y}$. However, Eq. (19) cannot be used for spectral CT directly since different energy bins have different intensities. Indeed, different energy-bin images always have different mean values. However, if the channel-wise spectral images can be normalized using their corresponding mean value, we can further normalize images to calculate SSIM. To address this issue, let $\mathcal{X}^o$ be the network intermediate output, $\mathbf{x}_i^o$ is i-th energy bin image of $\mathcal{X}^o$, and $\mathbf{x}_{i-norm}^o$ represents the normalization of $\mathbf{x}_i^o$. Let $\mathbf{x}_{ref}$ be normalized image from full-spectrum average image. Then we can express the $L_{ss}$ as

$$L_{ss} = 1 - \frac{1}{I}\sum_{i=1}^{I} SSIM(\mathbf{x}_{i-norm}^o, \mathbf{x}_{ref}). \quad (20)$$

Regarding the normalization, the following formula is used in this study
$$\mathbf{x}_{i-norm}^o = \frac{\mathbf{x}_i - min(\mathbf{x}_i)}{max(\mathbf{x}_i) - min(\mathbf{x}_i)}, \quad (21)$$
where $max(\cdot)$ represents the maximum value and $min(\cdot)$ represents the minimum value. The loss is given as
$$\underset{\theta}{argmin}(E_{\mathbf{s},\mathbf{x}}\|f_\theta(S_1(\mathbf{x})) - S_2(\mathbf{x})\|_F^2 + \lambda_s L_{ss} + \lambda_r L_r). \quad (22)$$
Eq. (22) shows that two regular terms $L_{ss}$ and $L_r$ play different roles in the process of network training. $L_r$ is good at image features recovery to avoid over smooth, and $L_{ss}$ benefits to speed-up the network fitting process and prevent detail losing. The overall flow of our network is shown in the Fig.1(b). For a given input image, we first down-sample the original images by employing a neighbor sub-sampler pair $S = (S_1, S_2)$ to generate a similar image pair $(S_1(\mathbf{x}), S_2(\mathbf{x}))$. Then, using $\mathbf{x}$ and $S_1(\mathbf{x})$ as the neural network inputs, we can further obtain the output $f_\theta(\mathbf{x})$ and $f_\theta((S_1(\mathbf{x}))$. Down-sampling $f_\theta(\mathbf{x})$ to generate $S_1(f_\theta(\mathbf{x}))$ and $S_2(f_\theta(\mathbf{x}))$, one can calculate the loss function using Eq. (22). As for the neural network, we employ the U-net to make full use of the characteristics of the image, and the architecture is shown in Fig.1(b). To construct the aforementioned down-sampler, Fig. 2 provides an intuitive diagram. For a given image $y$, it is divided into several $2 \times 2$ image patches firstly. For each image patch, two neighboring pixels are randomly chosen. For example, the pixel marked in yellow is taken as a pixel of a down-sampled image $S_1(\mathbf{y})$, and the other pixel marked in cyan is taken as a pixel of another down-sampled image $S_2(\mathbf{y})$.

## IV. EXPERIMENTS AND RESULTS

In this section, we first introduce the spectral CT dataset for reconstruction, including data preparation and neural network configuration. To validate our proposed unsupervised neural network for reconstruction, the classical FBP and conventional iteration reconstruction incorporating TV prior are selected for comparison. In addition, two post-processing methods (Noise2Sim and Noise2Noise) are implemented. To make a fair evaluation, these two methods are used as prior in an iterative framework. Since it is difficult to obtain clean labels, we only apply unsupervised deep learning algorithms for high-quality reconstruction. Noise2Sim has shown its capability in spectral CT reconstruction, and Noise2Noise has shown to be successful in reducing natural image noise. Additionally, SISTER (spectral-Image similarity-based tensor with enhanced-sparsity reconstruction) is also included for comparison [39]. Experimental results on the physical phantom and two preclinical mice demonstrate the advantages of our proposed method. The value of $\lambda_s$ is $e_c/e$, where $e_c$ is the current number of epoch and $e$ is total number of epoch. The value of $\lambda_r$ is 1. All the experiments use the same value of $\lambda_s$ and $\lambda_r$.

## A. Spectral Dataset and Network Configuration

### 1) Physical Phantom Data

To compare the performance of all the reconstruction methods, a home-made physical phantom is scanned by a spectral CT system (see Fig.3(a)).

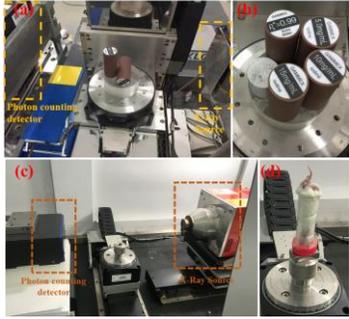

Fig. 3 The spectral CT imaging system and experiment materials. (a) is the spectral CT imaging system based on photon counting detector for physical phantom study, (b) is the physical phantom, (c) is another spectral CT imaging system for preclinical study, and (d) is the mice specimen.

The physical phantom consisting of three different basis materials (aluminum, water, and iodine). It contains five cylinders and their details can also refer to Fig.3(b). The spectral CT system has a 225-kV microfocus X-ray source from YXLON and PCD (XC-Hydra FX20) from Xcounter with two energy bins, more details refer to Fig.3(a). Here, the PCD contains 2048×8 detector units with each of detector pixel covering an area of 0.1×0.1 mm$^2$. In addition, every four cells along transverse direction are combined to improve SNR. The projection size is 512×180×5×8 with multiple scans, where 180 is the number of projection views. The process of scanning consumes 36 minutes. The distances starting from the X-ray source to the rotation axis and PCD are 182.68 mm and 440.50 mm, respectively, resulting in an FOV with a radius of 41.3 mm. The reconstructed material image includes 512×512 pixels each of which covers an area of 0.162×0.162 mm$^2$.

### 2) Preclinical Mice Data

To obtain preclinical dataset, a spectral CT imaging system based on a photon counting detector is employed to scan mice in our study as shown in Fig.3(c), and a mouse is shown in Fig.3(d). The photon-counting detector (the ME series of SANTIS 0804) manufactured by DECTRIS contains four fully independently adjustable energy gating thresholds. It consists of 515×257 pixels with each of detector pixel covering an area of 150×150 μm$^2$. The SANTIS 0804 is a Hybrid Photon Counting (HPC) detector, and it can configure high-resolution mode and multi-energy mode. The multi-energy mode is set to collect the projections, where the maximum tube voltage is 160kVp, and the maximum input photon counting rate is 4.0×10$^8$ photons/s/mm$^2$.

The scanned mice have been injected with relevant reagents to keep it sleeping during the scanning process. Their weights and lengths are roughly 160-180g and 10 cm. The distance starting from x-ray source to the detector and object are 350mm and 210mm, respectively. The emitting x-ray spectrum with 70kVp is divided into five energy bins: [20, 30)Kev, [30, 40)Kev, [40, 50)Kev, [50, 60)Kev and [60, 70)Kev. There are 250 projection angles uniformly distributed over 360°. For fan-beam geometry of the central slice, the projection size is 512×250×5, where 250 is the number of projection views.

### 3) Network configuration

The programming language used in the experiments is Python 3.6, the U-Net network is implemented based on the deep learning framework Pytorch 1.7.0, and the implementation of SIRT is based on the Astra Toolbox 1.9.0dev. The post-processing method is to denoise the result of SIRT reconstruction, where the number of SIRT iterations is 50. The hardware platform configuration for running each network is as follows: Intel i5-9600kf CPU, Nvidia TITAN V (12GB / Nvidia) GPU, and 16G DDR4 RAM. For each network training, the number of epochs is set to 50, the learning rate is $1.0\times10^{-4}$ and decreased to 90% per 10 epochs, and the Adam method is employed to optimize the network.

## B. Experimental Results

### 1) Physical Phantom

The x-ray energy spectrum range in this study is [20, 70]Kev. Multiple measurements are performed using the detector to obtain the photon numbers in the ranges of [20,70], [30, 70], [40, 70], [50, 70], and [60, 70], respectively. Then, post-processing step (subtraction) is applied to obtain the photon numbers on the energy bin [20, 30) and others. Fig.4 shows the reconstruction results using different reconstruction algorithms from the physical phantom. In our experiments, the parameters are expirically optimized for each method. When dealing with multiple parameters, all the parameters are alternatingly optimized. When one parameter is optimizing, other parameters are unchanged. For each parameter, we explore a range of values, and the optimal one is selected. The procedure is repeated till all the paramters are stable. From the 1$^{st}$ column in Fig.4, it can be seen that the FBP method provides the worst reconstruction results. These images are corrupted by noises since there are only a small number of photons within each energy bin. Although Noise2sim removes part of the noises, it generate obviously blurred images and the image quality is still insufficient to satisfy preclinical requirements. This is because there are only 8 slices in the physical phantom, and the Noise2sim needs as many different slices as possible to improve image denoising results. Compared with the Noise2sim, Noise2noise and SISTER can reconstruct relative clear images. SISTER method is better than Noise2Noise in terms of detail structures. Compared with the above methods, our proposed Spectral2Spectral has an ability to provide clear edges and structures. Indeed, our proposed reconstruction method significantly improves the image quality and further shows the effectiveness of Spectral2Spectral in spectral CT reconstruction.

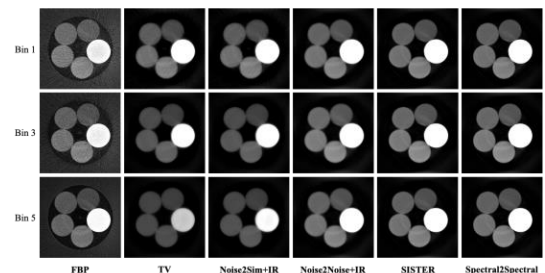

Fig. 4 Reconstruction results of the physical phantom. The 1$^{st}$-5$^{th}$ columns represent FBP, Noise2Sim+IR, Noise2Noise+IR, SISTER and Spectral2Spectral results. The 1$^{st}$ -3$^{rd}$ rows represent reconstructed results from 1$^{st}$, 3$^{rd}$, and 5$^{th}$ energy bins. The display window is [0, 0.015], [0,0.01].

*2) Material decomposition*

To further evaluate the reconstructed attenuation coefficients of the scanned object for each energy bin, we implement a material decomposition experiment. There are five phantoms used in this part. They are aluminum, solid water, 5mg/mL iodine, 10mg/mL iodine and 15mg/mL iodine, respectively. The ideal material maps can be theoretically computed to serve as references. For example, the ideal Aluminum map is a uniform disk with a value of 1.0. The results are shown in Fig.5. To establish a ground truth for quantitative evaluation, we computed the the ideal material decomposition results based on the known Iodine contcentrations. The ideal values are used as reference to calculate the root mean square error (RMSE), and the results are summarized in Table 1.

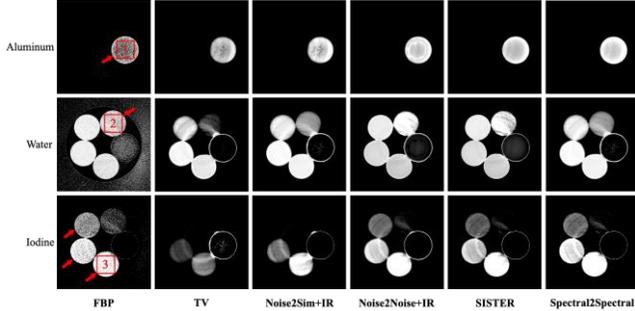

Fig. 5 Material decomposition results of the physical phantom. The $1^{st}$ -$3^{rd}$ rows represent Aluminum, water and Iodine, which are indicated by red arrows. The $1^{st}$ -$5^{th}$ columns represent different reconstruction methods. The display windows for the 1st-3rd rows are [0.5, 1], [0.75,1], [0,0.002].

Table 1. Quantitative RMSE evaluation of ROIs 1-3.

| ROI number | ROI-1 | ROI-2 | ROI-3 |
| --- | --- | --- | --- |
| FBP | $96.7 \times 10^{-3}$ | $139.2 \times 10^{-3}$ | $5.6 \times 10^{-3}$ |
| TV | $83.2 \times 10^{-3}$ | $120.1 \times 10^{-3}$ | $5.7 \times 10^{-3}$ |
| Noise2Sim+IR | $68.7 \times 10^{-3}$ | $100.5 \times 10^{-3}$ | $3.1 \times 10^{-3}$ |
| Noise2Noise+IR | $74.7 \times 10^{-3}$ | $123.3 \times 10^{-3}$ | $1.8 \times 10^{-3}$ |
| SISTER | $52.5 \times 10^{-3}$ | $112.6 \times 10^{-3}$ | $1.5 \times 10^{-3}$ |
| Spectral2Spectral | $50.2 \times 10^{-3}$ | $98.2 \times 10^{-3}$ | $1.3 \times 10^{-3}$ |

*3) Preclinical Mice*

To demonstrate the excellent reconstruction performance of Spectral2Spectral, Figs.6 and 7 show the reconstruction results with five energy bins of two different slices from two mice, respectively. Here, the extracted ROIs are also highlighted. The $1^{st}$ -$3^{rd}$ rows represent three different energy bins of 25-35Kev, 45-55Kev, and 65-75Kev, respectively. It can be seen from Figs. 6 and 7 that the images reconstructed using FBP have the strongest noise. Moreover, the image details in the $5^{th}$ energy bin have been deteriorated, and bone features are difficult to be distinguished. Noise2Sim and Noise2Noise can generate images almost as good as SISTER and Spectral2Spectra within bin 1. However, the first two methods show poor performance within bins 3 and 5. They reconstruct blurry images and lose detailed information.

Noise2Sim and Noise2Noise work well in non-edge places without grayscale abrupt changes. Benefiting from the iterative framework, the results of these two methods don't show obvious noises. The images are too smooth, especially within the $5^{th}$ bin. The Noise2Sim adopts the spatial similarity between adjacent layers of the image to be reconstructed, and it does not use the similarity between different energy bins along the spectral dimension. Furthermore, Noise2Sim performs well when the noise levels of all energy channels are consistent. And the distance between adjacent slices must close enough to avoid significant structural changes. The Noise2Noise method requires that noise is expected zero-mean. Because the $5^{th}$ bin has the most inherent noise, this method shows the worst performance for bin 5. SISTER and Spectral2Spectral methods can reconstruct clear images. But there are many weight coefficients in SISTER algorithm, which increase difficulty in algorithm implementation. Our Spectral2Spectral method don't have these limitations and obtain the best reconstruction results.

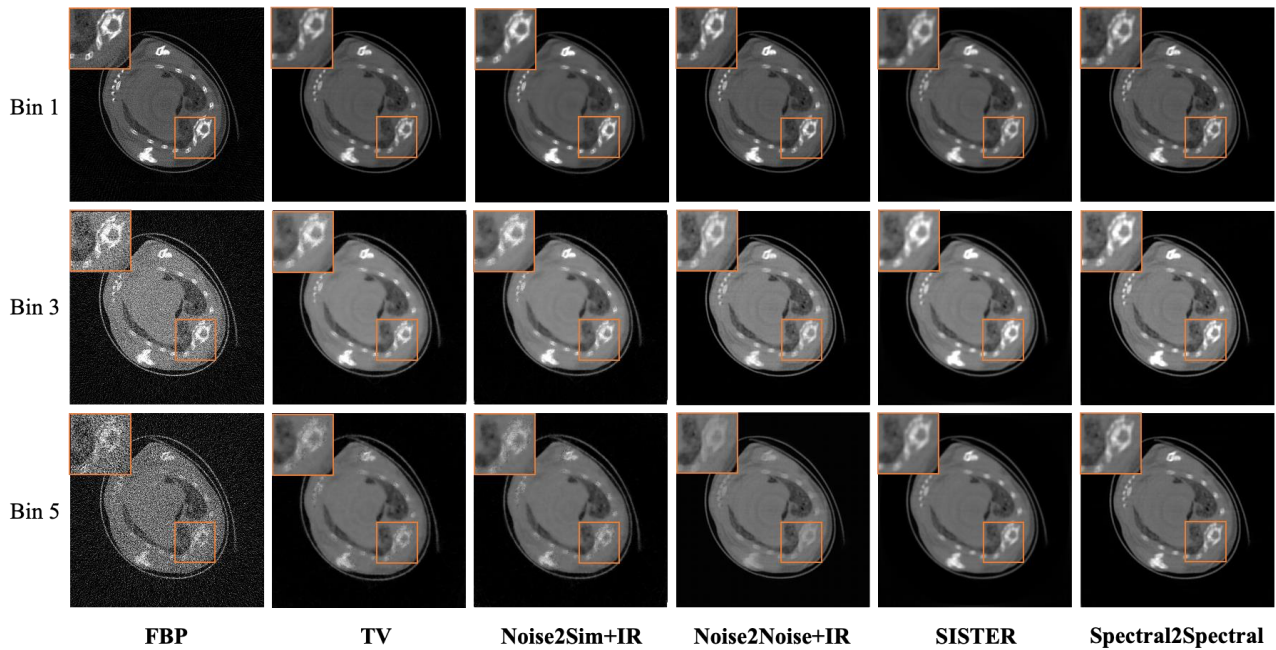

Fig. 6 Mouse reconstruction results of case 1 using different reconstruction algorithm. ROI is indicated by a box. From 1st-3rd row, the display windows are [0, 0.007], [0, 0.003], and [0, 0.003].

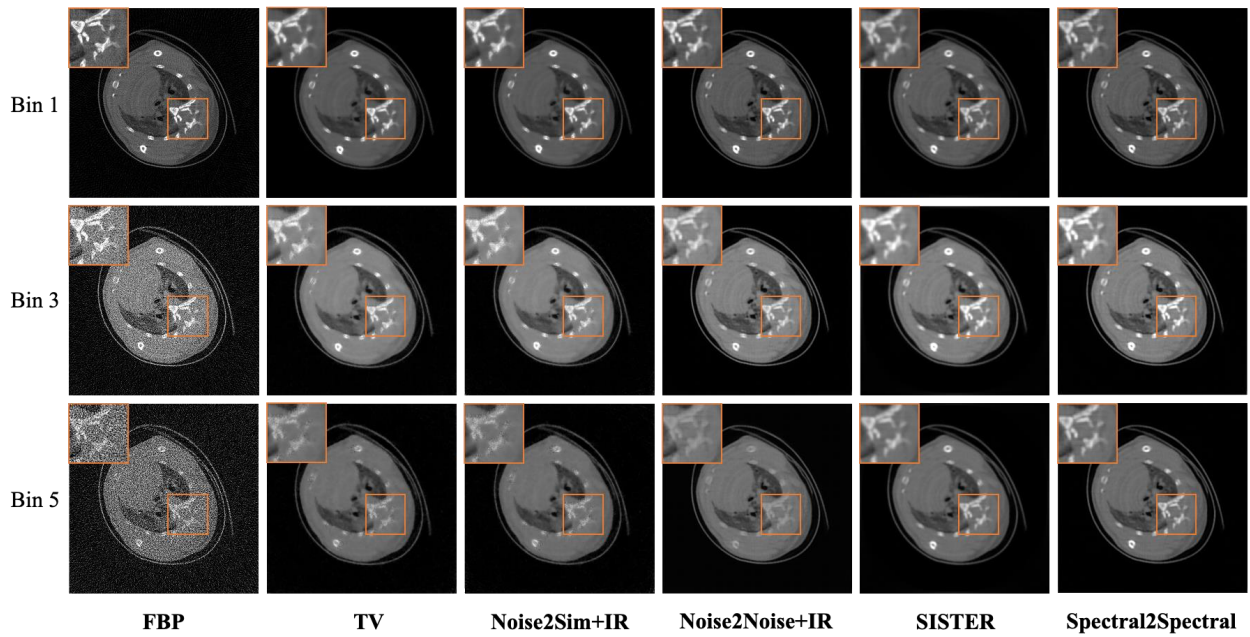

Fig. 7 Mouse reconstruction results of case 2 using different reconstruction algorithms. ROI is indicated by a box. From 1st-3rd row, the display windows are [0,0.007], [0,0.003], and [0,0.003].

Fig.8 and 9 also provide reconstruction results in terms of sagittal and coronal views. It can be seen from the red arrows in the 5[th] energy-bin images in Figs. 8 and 9, Spectral2Spectral provides the clearest bony structures. The coronal reconstructions shown in Fig. 8 are also exciting. For the reconstructed results in bin 3 and bin 5, FBP and Noise2Sim fail to clearly reconstruct the structures indicated by red arrows. Particularly, the bony structures in 5[th] energy-channel image are missing.

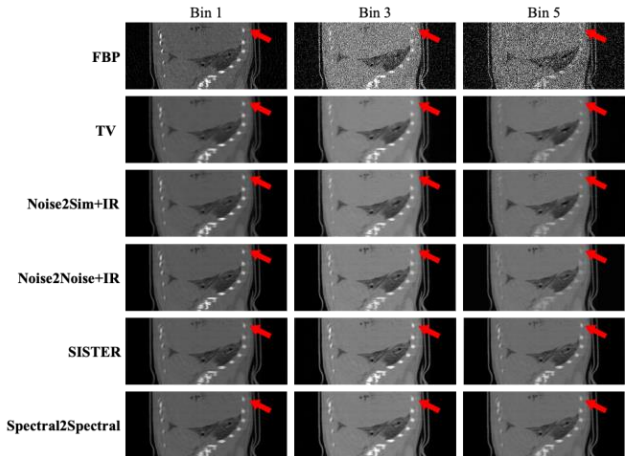

Fig. 8 The reconstructed results of one representative sagittal slice from mice. The display windows are set the same as Fig. 6.

From Fig.9, we can see that our method provides the best image quality, which is reflected in the following two aspects. First, the contrast resolution of the image is the highest. For example, there are blurred edges for Noise2Sim and Noise2Noise methods. In addition, these two methods even lose structure information of bone. On the contrary, the bone edges in our reconstruction results are clear. Second, different components can be clearly distinguished from each other. Compared with other competing methods, our

Spectral2Spectral provides the clearest image edge and structures as well as the highest contrast. The profile's results in Fig. 10 further demonstrate our proposed method obtains the best reconstruction results.

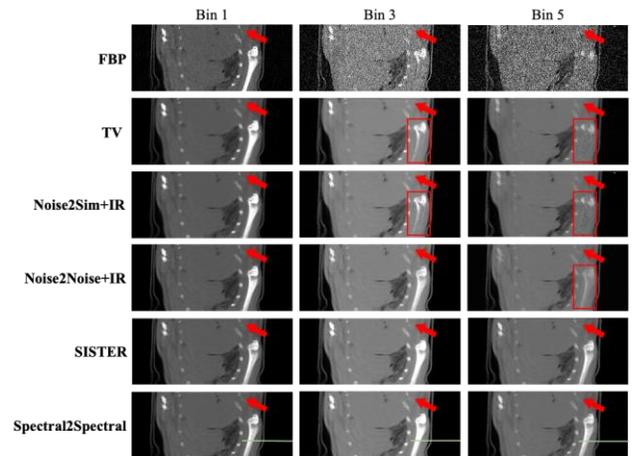

Fig. 9 The reconstructed results of one representative coronal slice from mice. The display windows are set as the same of Fig.6.

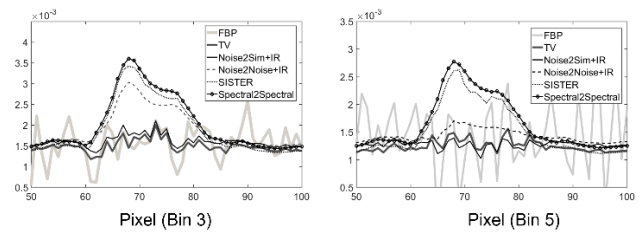

Fig. 10 Horizontal profiles through bone (green line) in images of Fig. 9

The visualization assessment demonstrates that the results are consistent with the quantitative evaluation results in Table 1. It can be seen that all the reconstruction methods can get a satisfying result when noise level is relatively low (bin 1). But for bin 3 and bin 5, most methods couldn't keep the same reconstruction performance. Compared with other reconstruction methods, our Spectral2Spectral has the best image quality in all the energy bins.

*C. Ablation study*

To evaluate the contributions of different modules in our proposed method, the ablation studies are performed. There are mainly three modules in the method, including unsupervised denoising network (DN), regularization optimization (RG), and iterative framework (IR). In Figs. 11 and 12, we show the reconstructed results of the model without DN, RG or IR, respectively. Without the DN module, the results have obvious noise. The model without RG can obtain almost the same results as the whole model within energy bin 1. However, for energy bins 3 and 5, it has poor image contrast resolution. For the model without IR, the reconstructed images are blurry within three energy bins.

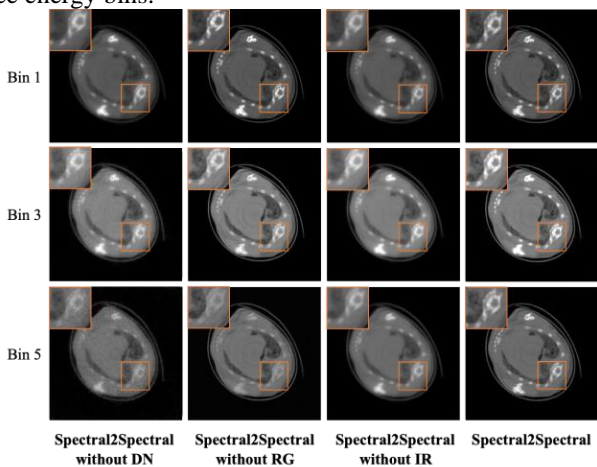

Fig. 11 Ablation study results for case 1. Results of $1^{st}$-$3^{rd}$ columns are reconstructed without DN, RG, or IR, respectively.

We also use SSIM to quantitatively evaluate the impacts of different modules. We choose the Specral2Spectral reconstructed images as references, the results are shown in Table 2. The DN module is mainly responsible for keeping structure of image and making image clean. The RG and IR modules can keep detail image structures and pixel values.

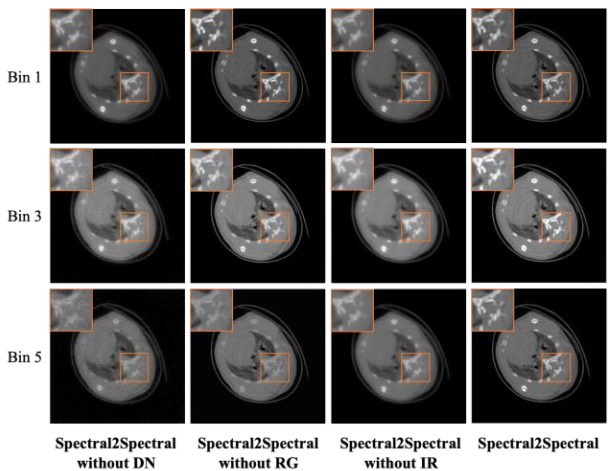

Fig. 12 Same as Fig. 11 but for case 2.

Table 2. Quantitative evaluation results of SSIM

| Bin number | 1 | 3 | 5 |
|---|---|---|---|
| Spectral2Spectral without DN | $0.9702 \pm 0.32\%$ | $0.9754 \pm 0.22\%$ | $0.9598 \pm 0.34\%$ |
| Spectral2Spectral without RG | $0.9980 \pm 0.01\%$ | $0.9985 \pm 0.01\%$ | $0.9899 \pm 0.04\%$ |
| Spectral2Spectral without IR | $0.9784 \pm 0.25\%$ | $0.9886 \pm 0.14\%$ | $0.9887 \pm 0.13\%$ |

## V. DISCUSSION AND CONCLUSION

According to our experimental results, our proposed deep iterative reconstruction algorithm, *i.e.,* Spectral2Spectral, can obtain high-quality images from spectral CT data with significantly improved reconstruction performance. Compared with CS-based reconstruction technique, Spectral2Spectral results provide high-quality images with less noise. Our method also outperforms the state-of-the-art neural network post-processing methods. We combine an iterative framework, unsupervised denoising network, and regularization optimization in a unified model. Although the traditional model-based iterative reconstruction can greatly improve image quality by increasing resolution as well as reducing noise and some artifacts, this type method has to suffer high computational cost and long reconstruction time. Our proposed method doesn't have these disadvantages and can obtain satisfactorily reconstruction results after the network is well-trained.

The network structure in our study is U-Net, which is not the latest network structure. CNN-like networks have strong local feature extraction capabilities, but these networks cannot capture global feature information. In the follow-up research, we can try to combine self-attention-based architectures like Transformers with CNN as a more powerful feature extractor [40, 41]. Second, we use the structural similarity among different energy channels to constrain the network training, without using some other prior information, such as the CT image sparsity and spatial similarity. Hence, more constraints can be added to further improve image quality. Third, we use the similarity between adjacent pixels to construct training pairs. Certainly, the extra regularization terms bring additional computation cost, and it may increase the bias of the model. In the further research, we can also explore the diffusion model based spectral CT reconstruction by modelling its data distribution [42].

In conclusion, we reconstruct spectral CT by combining deep learning with iterative reconstruction methods, and the reconstruction performance on narrow energy channels has been verified. It further demonstrates that deep learning can coexist with traditional reconstruction methods to obtain better image quality. It is believed that our proposed method has great potential in spectral CT reconstruction.

## REFERENCES

1. Donoho, D.L., *Compressed sensing.* IEEE Transactions on information theory, 2006. **52**(4): p. 1289-1306.
2. Teyfouri, N., H. Rabbani, and I. Jabbari, *Low-dose cone-beam computed tomography reconstruction through a fast three-dimensional compressed sensing method based on the three-dimensional pseudo-polar fourier transform.* Journal of Medical Signals and Sensors, 2022. **12**(1): p. 8.

3. Zhang, Y., et al., *Tensor-based dictionary learning for spectral CT reconstruction.* IEEE transactions on medical imaging, 2016. **36**(1): p. 142-154.
4. Wu, W., et al., *Low-dose spectral CT reconstruction using image gradient ℓ0–norm and tensor dictionary.* Applied Mathematical Modelling, 2018. **63**: p. 538-557.
5. Cueva, E., et al., *Synergistic multi-spectral CT reconstruction with directional total variation.* Philosophical Transactions of the Royal Society A, 2021. **379**(2204): p. 20200198.
6. Kong, H., et al., *Spectral CT reconstruction based on PICCS and dictionary learning.* IEEE Access, 2020. **8**: p. 133367-133376.
7. Wang, Q., et al., *Locally linear transform based three‐dimensional gradient‐norm minimization for spectral CT reconstruction.* Medical Physics, 2020. **47**(10): p. 4810-4826.
8. Yu, Z., et al., *Spectral prior image constrained compressed sensing (spectral PICCS) for photon-counting computed tomography.* Physics in Medicine & Biology, 2016. **61**(18): p. 6707.
9. He, Y., et al., *Spectral CT reconstruction via low-rank representation and structure preserving regularization.* Physics in Medicine & Biology, 2023. **68**(2): p. 025011.
10. Shi, Y., et al., *Spectral CT reconstruction via low-rank representation and region-specific texture preserving Markov random field regularization.* IEEE transactions on medical imaging, 2020. **39**(10): p. 2996-3007.
11. Chen, X., et al. *Fourth-order nonlocal tensor decomposition model for spectral computed tomography.* in *2021 IEEE 18th International Symposium on Biomedical Imaging (ISBI).* 2021. IEEE.
12. Rigie, D.S. and P.J. La Riviere, *Joint reconstruction of multi-channel, spectral CT data via constrained total nuclear variation minimization.* Physics in Medicine & Biology, 2015. **60**(5): p. 1741.
13. Zhou, Z., *Direct iterative basis image reconstruction based on MAP-EM algorithm for spectral CT*, in *Advanced X-Ray Radiation Detection: Medical Imaging and Industrial Applications.* 2022, Springer. p. 219-238.
14. Chen, H., et al. *Low-dose CT denoising with convolutional neural network.* in *2017 IEEE 14th International Symposium on Biomedical Imaging (ISBI 2017).* 2017. IEEE.
15. Chen, H., et al., *Low-dose CT with a residual encoder-decoder convolutional neural network.* IEEE transactions on medical imaging, 2017. **36**(12): p. 2524-2535.
16. Jin, K.H., et al., *Deep convolutional neural network for inverse problems in imaging.* IEEE Transactions on Image Processing, 2017. **26**(9): p. 4509-4522.
17. Ronneberger, O., P. Fischer, and T. Brox. *U-net: Convolutional networks for biomedical image segmentation.* in *Medical Image Computing and Computer-Assisted Intervention–MICCAI 2015: 18th International Conference, Munich, Germany, October 5-9, 2015, Proceedings, Part III 18.* 2015. Springer.
18. Zhang, Z., et al., *A sparse-view CT reconstruction method based on combination of DenseNet and deconvolution.* IEEE transactions on medical imaging, 2018. **37**(6): p. 1407-1417.
19. Kang, E., J. Min, and J.C. Ye, *A deep convolutional neural network using directional wavelets for low‐dose X‐ray CT reconstruction.* Medical Physics, 2017. **44**(10): p. e360-e375.
20. Han, Y. and J.C. Ye, *Framing U-Net via deep convolutional framelets: Application to sparse-view CT.* IEEE transactions on medical imaging, 2018. **37**(6): p. 1418-1429.
21. Chun, Y. and J.A. Fessler. *Deep BCD-net using identical encoding-decoding CNN structures for iterative image recovery.* in *2018 IEEE 13th Image, Video, and Multidimensional Signal Processing Workshop (IVMSP).* 2018. IEEE.
22. Chun, I.Y., et al., *Momentum-Net: Fast and convergent iterative neural network for inverse problems.* IEEE Transactions on Pattern Analysis and Machine Intelligence, 2020.
23. He, J., Y. Wang, and J. Ma, *Radon inversion via deep learning.* IEEE transactions on medical imaging, 2020. **39**(6): p. 2076-2087.
24. Zhu, B., et al., *Image reconstruction by domain-transform manifold learning.* Nature, 2018. **555**(7697): p. 487-492.
25. Niu, C., et al., *Noise suppression with similarity-based self-supervised deep learning.* IEEE Transactions on Medical Imaging, 2022.
26. Lehtinen, J., et al., *Noise2Noise: Learning image restoration without clean data.* arXiv preprint arXiv:1803.04189, 2018.
27. Krull, A., T.-O. Buchholz, and F. Jug. *Noise2void-learning denoising from single noisy images.* in *Proceedings of the IEEE/CVF conference on computer vision and pattern recognition.* 2019.
28. Batson, J. and L. Royer. *Noise2self: Blind denoising by self-supervision.* in *International Conference on Machine Learning.* 2019. PMLR.
29. Quan, Y., et al. *Self2self with dropout: Learning self-supervised denoising from single image.* in *Proceedings of the IEEE/CVF conference on computer vision and pattern recognition.* 2020.
30. Pang, T., et al. *Recorrupted-to-recorrupted: Unsupervised deep learning for image denoising.* in *Proceedings of the IEEE/CVF conference on computer vision and pattern recognition.* 2021.
31. Wu, W., et al., *Deep learning based spectral CT imaging.* Neural Networks, 2021. **144**: p. 342-358.
32. Sidky, E.Y. and X. Pan, *Image reconstruction in circular cone-beam computed tomography by constrained, total-variation minimization.* Physics in Medicine & Biology, 2008. **53**(17): p. 4777.
33. Gao, H., et al., *Multi-energy CT based on a prior rank, intensity and sparsity model (PRISM).* Inverse problems, 2011. **27**(11): p. 115012.
34. Zhang, Y., et al., *Spectral CT reconstruction with image sparsity and spectral mean.* IEEE transactions on computational imaging, 2016. **2**(4): p. 510-523.
35. Sidky, E.Y., C.-M. Kao, and X. Pan, *Accurate image reconstruction from few-views and limited-angle data in divergent-beam CT.* Journal of X-ray Science and Technology, 2006. **14**(2): p. 119-139.
36. Qiao, Z., et al., *A Simple but Universal Fully Linearized ADMM Algorithm for Optimization Based Image Reconstruction.* 2023.
37. Huang, T., et al. *Neighbor2neighbor: Self-supervised denoising from single noisy images.* in *Proceedings of the IEEE/CVF conference on computer vision and pattern recognition.* 2021.
38. Wang, M., et al., *An adaptive reconstruction algorithm for spectral CT regularized by a reference image.* Physics in Medicine & Biology, 2016. **61**(24): p. 8699.
39. Hu, D., et al., *SISTER: Spectral-image similarity-based tensor with enhanced-sparsity reconstruction for sparse-view multi-energy CT.* IEEE Transactions on Computational Imaging, 2019. **6**: p. 477-490.
40. Wu, W., et al., *DRONE: Dual-domain residual-based optimization network for sparse-view CT reconstruction.* IEEE Transactions on Medical Imaging, 2021. **40**(11): p. 3002-3014.
41. Wu, W., et al., *Deep embedding-attention-refinement for sparse-view CT reconstruction.* IEEE Transactions on Instrumentation and Measurement, 2022. **72**: p. 1-11.
42. Wu, W., et al., *Wavelet-improved Score-based Generative Model for Medical Imaging.* IEEE Transactions on Medical Imaging, 2023.

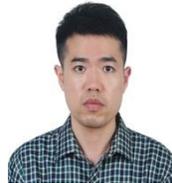

**Xiaodong Guo** received the B.S. degree from the University of Electronic Science and Technology of China, Sichuan, China, in 2014, and the M.Sc. degree from Southwest Jiaotong University in 2018. He is currently pursuing the Ph.D. degree in optical engineering with Chongqing University, Chongqing, China. He is also currently doing cooperation research at the Rensselaer Polytechnic Institute, Troy, NY, USA. His research interests include the application of deep learning in medical CT images and image processing.

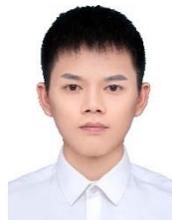

**Yonghui Li** received the B.S. degree from the Chongqing University, Chongqing, China, in 2020. He is currently pursuing the M.Sc. degree in instrumentation science and technology with Chongqing University, Chongqing, China, under the supervision of Prof. Peng He. His research interests include the application of deep learning in medical CT images and image processing.

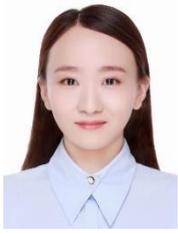
**Dingyue Chang** received the Ph.D. degree from the School of Earth and Space Science, Peking University, Beijing, China, in 2019. Since 2019, she has been an Engineer with the Institute of Materials, China Academy of Engineering Physics, Mianyang, China, where she is currently a Manager of the Laboratory on Microfocus Computed Tomography (CT). Her research interests include digital signal processing for microfocus CT systems and ultrasonic systems, novel imaging methods, and construction of ultrasound research systems.

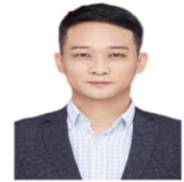
**Peng He** received the B.S. degree from Nanchang Hangkong University, Nanchang, Jiangxi, China, in 2007, and the Ph.D. degree in optical engineering from Chongqing University, Chongqing, China, in 2013. He is currently pursuing the joint Ph.D. degree in biomedical engineering with Virginia Polytechnic Institute and State University. From July 2013 to June 2015, he held a postdoctoral position with Chongqing University. From July 2015 to December 2016, he was a Lecturer of instrument science and technology with Chongqing University, where he is currently an Associate Professor with the Department of Optoelectronics Engineering. His research interests include X-ray spectral CT imaging, digital image processing, and big data artificial intelligence. He has published more than 30 articles in his research areas.

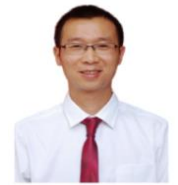
**Peng Feng** received the B.S. degree in mechanical and electronics engineering and the Ph.D. degree in optics engineering from Chongqing University, in 2002 and 2007, respectively, which is one of the most prestigious universities in China. From June 2008 to January 2012, he was a Postdoctoral Fellow with the School of Biomedical Engineering, Chongqing University, where he is currently an Associate Professor with the Department of Optoelectronics Engineering. He has published more than 30 peer-reviewed journal articles. His interests include computed tomography, compressive sensing, image representation, and biomedical image processing.

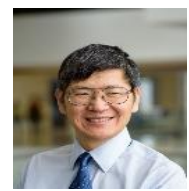
**Hengyong Yu(SM'06)** received his Bachelor degrees in information science & technology and computational mathematics in 1998, and his PhD degree in Electrical Engineering in 2003 from Xi'an Jiaotong University. Currently, his research focuses on data-driven artificial intelligence methods for medical imaging and image processing, with an emphasis on computed tomography. He has authored/coauthored >200 peer-reviewed journal papers with an H-index of 46. As PI or key investigator, he has received >20 major grants with a total budget of >$25M. He was the founding Editor-in-Chief of *JSM Biomedical Imaging Data Papers*, serves as an Editorial Board member or associate editor for *IEEE Transactions on Medical Imaging, Medical Physics, IEEE Access*, *Signal Processing*, *etc*. In January 2012, he received an NSF CAREER award for development of CS-based interior tomography. In September 2022, he received the IEEE R1 Technological Innovation Award (Academic) for "*pioneering contributions and international leadership in tomographic imaging, especially interior tomography and machine learning-based tomographic imaging*".

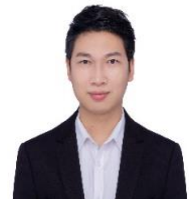
**Weiwen Wu** received the Ph.D degree in Instrument Science and Technology from Chongqing University, Chongqing, China, in 2019. From 2019 to 2021, he completed his postdoctoral training in Hong Kong University and Renssenlear Polytechnique Institute under supervision of Professor Ge Wang. He is an associate professor now with the school of biomedical engineering, Sun Yat-sen University of China. His research interests focus on computed tomography, including imaging system design, compressed sensing-based and deep learning-based image reconstruction.